\documentclass[aps,twocolumn,prl]{revtex4}
\usepackage{graphicx}
\usepackage{amsfonts}

\begin{document}

\draft

\title{Atomic Gases at Negative Kinetic Temperature}

\author{A. P. Mosk}
\affiliation{Department of Science \& Technology, and MESA+
Research Institute, University of Twente, P.O. Box 217, 7500 AE
Enschede,
The Netherlands\\
and Ecole Normale Sup\'{e}rieure, Laboratoire Kastler-Brossel, 24 rue Lhomond, 75231 Paris CEDEX 05, France}
\date{Version 3, June 8, 2005}
\begin{abstract}
We show that thermalization of the motion of atoms at negative
temperature is possible in an optical lattice, for conditions that
are feasible in current experiments. We present a method for
reversibly inverting the temperature of a trapped gas. Moreover, a
negative-temperature ensemble can be cooled (reducing $|T|$) by
evaporation of the lowest-energy particles. This enables the
attainment of the Bose-Einstein condensation phase transition at
negative temperature.
\end{abstract}
 \maketitle

Amplification of a macroscopic degree of freedom, such as a laser
or maser field, or the motion of an object, is a phenomenon of
enormous technological and scientific importance. In classical
thermodynamics, amplification is not possible in thermal
equilibrium at positive temperature. In contrast, due to the
fluctuation-dissipation theorems \cite{Einstein1905,Kubo1966},
amplification is a natural phenomenon, and fully consistent with
the second law of thermodynamics, in a negative-temperature heat
bath \cite{Strandberg1957,Geusic1967,Nakagomi1980,Hsu1992}.
Negative temperatures were introduced by Purcell and Pound
\cite{Purcell1951} to describe spin systems, and further
generalized theoretically by Ramsey and Klein
\cite{Ramsey1956,Klein1956}. Useful introductions to
thermodynamics at negative temperature are given by Landau and
Lifshitz \cite{LandauLifshitz} and by Kittel and Kroemer
\cite{KittelKroemer}.
 Negative temperatures are used to
describe distributions where the occupation probability of a
quantum state increases with the energy of the state. The
prerequisite for a negative-temperature ensemble to be in internal
equilibrium is that it occupies a part of phase space where the
entropy decreases with internal energy. An ensemble in this part
of phase space cannot be in thermal equilibrium with any
positive-temperature system.
 An increasing occupation number (and
a decreasing entropy) can only exist up to a certain maximum
energy. Unlike the magnetic energy of spins, kinetic energy is not
bounded from above in free space, which apparently rules out a
negative kinetic temperature. Here, we show that the band
structure of a moderately deep optical lattice defines an upper
bound to the kinetic energy of atoms in the first band. Therefore,
atoms in an optical lattice
 form a simple system in which thermalization of the kinetic
energy is possible at negative temperatures.

Optical lattices, which are periodic potentials created through
optical standing waves \cite{Jessen1996}, impose a band structure
on the motion of ultracold atoms. The kinetic energy of atoms in
the lattice depends on their effective mass or band-mass, which,
in analogy to the case of electrons in a semiconductor, can be
strongly different from the normal rest mass. The effective mass
can even become negative, as has been demonstrated in
one-dimensional optical lattices \cite{Dahan1996,Eiermann2003}. An
isotropic negative effective mass can be realized in the vicinity
of a band gap in a three-dimensional (3D) optical lattice. The
bandgap is an energy range inside which no
 single-particle eigenstates exist.  If the atoms are constrained
 to energies just below the bandgap the effective mass will be negative in all directions and
 the entropy of the
 ensemble will decrease with energy. This is the condition for
 existence of a thermal ensemble with negative temperature \cite{Ramsey1956}.

For thermalization to take place, the characteristic timescale for
exchange of energy between particles must be shorter than
timescales associated to heating and loss. In other terms, we need
a high rate of intraband scattering, with negligible interband
scattering. Interband scattering is a collision process between
two atoms in the first (``valence'') band which causes one atom to
be promoted to the second (``conduction'') band while the
``valence'' band atom loses energy. Additional losses may be
caused by interband tunneling, a one-body process in which an atom
tunnels through a classically forbidden region, to re-appear in a
different band.

 In this Letter, we show that all the requirements for
 thermalization at negative temperature can be met in a 3D optical
 lattice. First, we show that an optical lattice can have a 3D bandgap for atoms, at experimentally
 feasible lattice intensities. Moreover, interband scattering can
 be completely suppressed, and confinement of the negative-mass
 atoms can be easily accomplished using a magnetic trap for
 {\it high-field seeking} states. We calculate rates of interband
 tunneling and find they can be made negligible. Finally, we show
 how a trapped gas at positive temperature can be reversibly converted to a
 negative temperature.

\begin{figure}
\center
\includegraphics[width=8 cm]{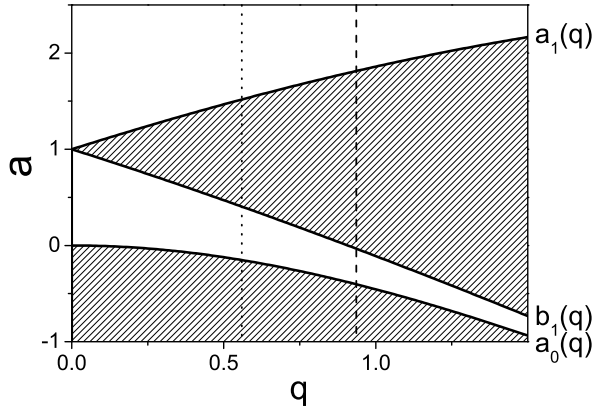}
\caption{Band structure of the one-dimensional Mathieu equation.
White area: Bands, grey area: gaps. Dotted line: $q=0.559$, dashed
line: $q=0.936$, see text.}
\label{fig:bands}       
\end{figure}

\begin{figure}
\center
\includegraphics[width=8 cm]{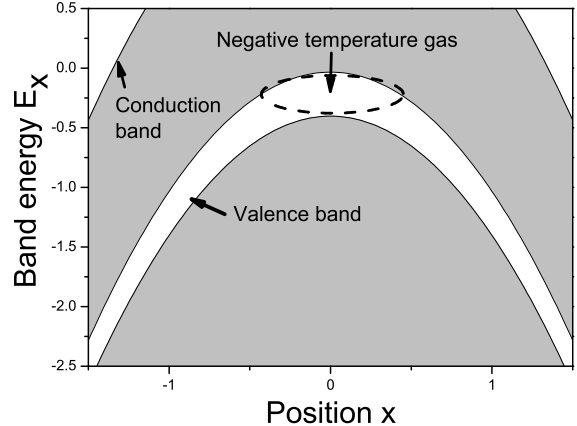}
\caption{``Valence'' and ``conduction'' bands of a 1D optical
lattice at $q=0.936$. Grey area: Gaps, white area: Bands.
   }
\label{fig:parabola}       
\end{figure}

 The motion of atoms in a simple cubic 3D optical lattice with a harmonic confining potential
 is
 described by the Schr\"odinger equation (See e.g., \cite{Jessen1996,Pu2003}),
 \begin{eqnarray}
 \label{Eq:schroedinger}
 E \Psi &=& \left[ \frac{-\hbar^2}{2m} \nabla^2 + U_{\rm lat}+
 \alpha (x^2+y^2+z^2) \right] \Psi \\ \nonumber
 {\rm with} && U_{\rm lat}= {U_0} (\cos^2 kx  + \cos^2 ky +
 \cos^2 kz ).
 \end{eqnarray}
Here  $m$ is the bare atomic mass, $\alpha$ is the strength of the
harmonic potential, $U_0$ is the depth of the standing wave
potentials that form the lattice, and $k= 2\pi/\lambda$, where
$\lambda$ is the wavelength of the lattice light. The above
Schr\"odinger equation is trivially separable by setting
$E=E_x+E_y+E_z$. We now transform the equation to dimensionless
units,
\begin{eqnarray}
 \label{Eq:subs}
\tilde{x},\tilde{y},\tilde{z}&=& kx,ky,kz \\
\tilde{\alpha}&=& \alpha/(E_r k^2) \\
a_x&=& (E_x- \frac{1}{2}U_0)/E_r \\
q &=& U_0/4E_r,
\end{eqnarray}
with the lattice recoil energy $E_r=\hbar^2 k^2/2m$. The
one-dimensional equation of motion takes the shape of a modified
Mathieu equation,
\begin{equation}
\label{Eq:Mathieu}
 \frac{d^2 \Psi}{d {\tilde{x}}^2} + (a_x  - 2 q \cos 2 \tilde{x}) \Psi + \tilde{\alpha} \tilde{x}^2 \Psi
 =0.
\end{equation}
In the absence of a parabolic potential ($\tilde{\alpha}=0$),
Eq.~(\ref{Eq:Mathieu}) is exactly the Mathieu equation. Its
solutions are the Mathieu functions, which are propagating
Bloch-Flouquet waves for values of $a$ and $q$ in the bands, and
evanescent functions in the bandgaps. The relevant bands are
depicted in Fig.~\ref{fig:bands}. Adopting the usual convention
\cite{Gradshteyn} we designate the eigenvalues corresponding to
the band bottoms by $a_0(q), a_1(q),$ etc., and the band tops by
$b_1 (q)$, etc. (see Fig. \ref{fig:bands}). If $0<
|\tilde{\alpha}| \ll q/\tilde{x}$, the lattice is locally only
weakly perturbed by the parabolic potential, and the band
structure will remain essentially intact. The energy positions of
the bands become position dependent, in the same way as in a
semiconductor in an applied electric field. A negative value of
$\tilde{\alpha}$ corresponds to a confining potential for
negative-effective mass atoms, see Fig.~\ref{fig:parabola}. The
area under the ``valence'' band in Fig.~\ref{fig:parabola} is
classically forbidden as the kinetic energy of the atoms in this
region is less than zero. The area between the bands is Bragg
forbidden, i.e., no propagating modes can exist here due to
interference.

For the negative-temperature phase to be metastable, there must be
a bandgap between the top of the ``valence'' band, where the
negative mass states exist, and the bottom of the ``conduction''
band, where the effective mass is positive. In a 1D system (with
the motion in the other two directions confined) this bandgap
exists for $|q|>0$. However, in a 3D optical lattice, the bands
may overlap. The maximum energy an atom in the ``valence'' band of
a 3D lattice can have is $3b_1(q)$, whereas the minimum energy in
the ``conduction'' band is $2a_0(q)+a_1(q)$. The criterium for the
existence of a bandgap is $3 b_1 (q) < 2a_0(q)+a_1(q)$, which is
fulfilled for $q > 0.559$.

If the 3D bandgap is narrow, collisions between two ``valence''
band atoms can promote an atom to the ``conduction'' band, where
it has positive effective mass. This interband scattering will not
only lead to loss of atoms, but also to loss of energy, which is
equivalent to heating (increasing $|T|$).  We show now that
interband scattering becomes energetically forbidden in two-body
processes at high enough $q$. The maximum energy of the initial
state, consisting of two ``valence'' band atoms, is $6b_1(q)$. The
minimum energy of the final state, consisting of one ``valence''
and one ``conduction'' band atom, is $[3
a_0(q)]+[2a_0(q)+a_1(q)]$. The interband scattering process is
energetically forbidden if $6b_1(q) < 5 a_0(q)+a_1(q)$, which is
the case if $q > 0.936$. This value of $q$ corresponds to a
lattice depth almost an order of magnitude less than the depth
employed to create a Mott insulator phase \cite{Bloch2002}. In
such a moderately deep lattice, thermalization rates and inelastic
collision rates are all of the same order of magnitude  as the
respective rates at positive temperature in the absence of a
lattice \cite{Pu2003}. Through elastic collisions,  the ensemble
will always relax to a distribution with the highest possible
entropy at its given energy, in this case a negative-temperature
distribution.

\begin{figure}
%
\includegraphics[width=7 cm]{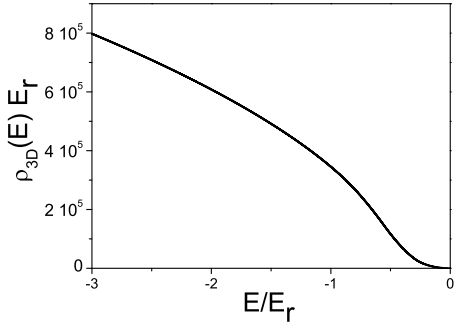} \caption{Density of
states $\rho(E)$ in a three-dimensional lattice with
$\bar{\alpha}=-10^{-3}$. The energy $E$ is taken with respect to
the top of the first (``valence'') band.
   }
\label{fig:dos}       
\end{figure}

To investigate how the gas thermalizes in the negative-temperature
phase space, it is useful to consider the semiclassical density of
states (DOS). The DOS can usually be found by taking the energy
derivative of the Wentzel-Kramers-Brillouin (WKB) phase integral,
a procedure recently applied to a one-dimensional optical lattice
with harmonic confinement (positive $\alpha$)
\cite{Hooley2004prl}. For negative $\alpha$ the WKB method cannot
be used straightforwardly to calculate wavefunctions, as the
``Bragg reflection'' boundary condition at the outer turning
points behaves differently from classical turning points. The WKB
formula for the DOS does not depend on the boundary conditions and
can be straightforwardly calculated in the one-band model of
\cite{Hooley2004prl}. Within the one-band and WKB approximations,
the DOS $\rho_{\alpha}$ at negative $\alpha$ is related to the DOS
calculated for positive $\alpha$ by
\begin{equation}
\rho_\alpha(E)=\rho_{-\alpha}(-E),
\end{equation}
where in the case of negative $\alpha$, the top of the valence
band is taken as the energy zero.
 The 3D DOS crosses over smoothly from a quadratic behavior
at small negative energy to approximately square-root behavior at
larger negative energy, as shown in Fig.~\ref{fig:dos}. The DOS is
nonzero for any energy lower than the band edge at the trap
center, so in principle the negative temperatures are \textit{not
bounded}\ in an infinite lattice. In fact, this asymmetric
behavior of the DOS makes thermal equilibrium at a positive
temperature impossible, which is in contrast to spin systems
\cite{Purcell1951}, where thermal equilibrium at positive and
negative temperature can be studied in the same field
configuration.

We now examine the escape of the atoms through interband (Zener)
tunneling. The trapped negative-mass ``valence'' band states are
mixed with anti-trapped ``conduction'' band states at the same
energy, as the evanescent tails of these states extend through the
Bragg forbidden region. The Zener tunneling rate $\Gamma_{\rm
zener}$  may be estimated using \cite{Zener1934},
\begin{equation}
 \Gamma_{\rm zener} \approx \omega_{\rm A}
 \exp\left[-2 \int_0^\infty {\rm Im}( k_{\rm Bloch}(x)) dx \right],
 \label{Eq:Zener}
\end{equation}
 where $k_{\rm Bloch}(x)$ is the Bloch wavevector of the Mathieu equation corresponding to the energy $E-\alpha x^2$,
and  $\omega_{\rm A}$ is an attempt frequency of the order of the
trap oscillation frequency (typically $< 1$ kHz). The resulting
tunneling time constant is of order $10^5$ seconds at
$\tilde{\alpha} = 10^{-3}$. For $10^{-4} < \tilde{\alpha}<
10^{-2}$ we checked
  Eq.~\ref{Eq:Zener} numerically by high-precision quadrature of
Eq.~\ref{Eq:Mathieu},
and found agreement within an order of magnitude.

\begin{figure}
\center
\includegraphics[width=8 cm]{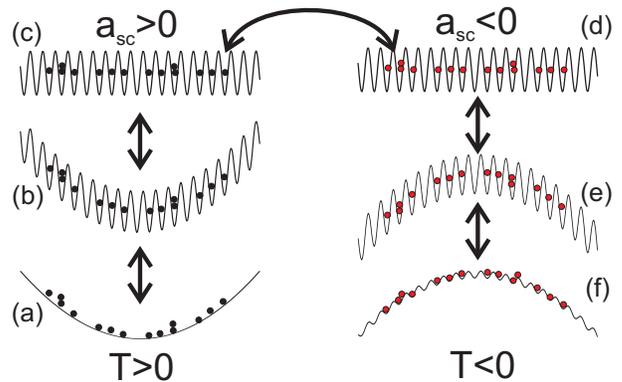}
\caption{Method for reversibly transforming a positive-temperature
gas (a) into a negative-temperature gas (f). Each step is
reversible, see text. (a) trapped gas at positive $T$. (b) a
strong optical lattice, (c) a strong optical lattice without trap,
(d) a strong optical lattice with changed $a_{sc}$, (e) a strong
optical lattice with negative trap, (f) an optical lattice at
$q=0.936$ with a negative-temperature gas.}
\label{fig:scheme}       
\end{figure}

The conditions for existence of a negative-mass Bose-Einstein
condensate (BEC) were recently studied by Pu and coworkers
\cite{Pu2003}, they also investigated the possibility of creating
such a condensate by phase imprinting, and found this likely to be
an inefficient mechanism as many atoms are transferred to other
bands. Other mechanisms, such as accelerating the atoms uniformly
to complete one-half Bloch oscillation, may be more efficient,
especially since in a BEC nonlinearity due to the interactions
counteracts dephasing \cite{Zheng2004, Brand2005}. Using these
methods, the negative-mass BEC can be created in a mechanically
unstable state as the interatomic scattering length $a_{sc}$
remains positive \cite{Pu2003}.

We propose a radically different mechanism for the production of
negative-temperature gases, based on the reversible creation of a
Mott-insulator phase from a BEC \cite{Bloch2002}. The idea is to
``freeze'' a positive temperature gas into a Mott insulator, by
turning on a strong optical lattice. The Mott phase is then
``molten'' in a different trap configuration and with different
interatomic interactions, to produce a negative-temperature gas.
The procedure is outlined in Fig.~\ref{fig:scheme}, we will
describe it here for a pure BEC but there is no {\it a priori}
reason why it would not work when some thermal excitations are
present.
 We start with a trapped BEC at $\alpha>0$ and
$a_{sc}>0$ (a), and turn on a strong optical lattice, as is done
in \cite{Bloch2002}. A Mott insulator phase will form in which the
atoms are confined to their local potential wells (b). We assume
one atom is present per lattice site. Now we remove the trap (c),
which can be done rapidly without creation of entropy, as the
atoms are tightly bound to their respective lattice sites. Then we
change the sign of $a_{sc}$ (c,d),
for example, by tuning the magnetic bias field near a Feshbach
resonance (See e.g., \cite{Inouye1998}). Since the atoms are at
different lattice sites and do not interact, this can be done
rapidly without causing losses. The system passes through the
point where both $a_{sc}$ and $\alpha$ are zero, here the
temperature of the Mott phase is not defined, but the entropy is.
This passage transforms the Mott insulator from the lowest-energy
many-particle state to the highest energy in the ``valence'' band.
 The confining potential is
reconfigured to change the sign of $\alpha$ (e). This can be done
either by changing the confinement field, or by changing the
internal state of the atoms. A magnetic field minimum will trap
negative-mass atoms in their lowest Zeeman state \cite{Pu2003}. No
change in the spatial distribution of atoms will take place due to
the extremely low tunneling rate in the Mott insulator, so this
step can proceed rapidly. When the lattice is slowly reduced in
strength (f), the Mott insulator will melt into high energy states
in the ``valence'' band, i.e., a negative-mass,
negative-temperature gas. This approach seems practicable for all
ultracold bosonic atoms which have suitably broad Feshbach
resonances, such as $^7$Li, $^{23}$Na, $^{85}$Rb, and $^{133}$Cs.
A detailed analysis of the reversibility criteria for each step
will be published elsewhere.

The atomic gas can be cooled by forced evaporation. In the case of
a magnetic trap, the evaporation can proceed through radio
frequency (RF) induced spin flips to an untrapped (magnetic
quantum number $m=0$) state. One should tune the RF to selectively
remove atoms with
 lower-than-average energy, so that
 the average energy per atom will increase. This leads to an increase of $T$,
 and a reduction of the entropy per particle $S/N$. A reduction of the entropy per particle
 is equivalent to an increase of the phase space density. Therefore we use the term {\em cooling}
 for any process that lowers $|T|$. At low enough $|T|$ the system can be described by a
(negative) power law density of states, and the theory of
evaporative cooling in a power law density of states
 \cite{Luiten1996} can be applied. (One should take care to reverse
the signs of $m$, $E$ and $T$ in Ref.\cite{Luiten1996} wherever
applicable.)
 The condition for BEC in a 3D system is $n \Lambda^3 = 2.61$, where
 $n$ is the number density of atoms and $\Lambda^2=2 \pi \hbar^2
 m_{\rm eff}^{-1} k_B^{-1}T^{-1}$, where $k_{B}$ is Boltzmann's constant. Since both the effective mass $m_{\rm eff}$ and
 $T$ are negative, $\Lambda$ is a real number.
  Achieving the BEC phase transition
 at negative temperature seems possible, either by heating
 (extracting energy), starting from a pure BEC, or by cooling,
 starting from a negative-thermal gas.

The negative value of $a_{sc}$, which is necessary to create the
negative-mass BEC from a Mott insulator, is also a condition of
stability for the negative-mass BEC. In fact, the negative-mass
BEC is likely to collapse if $a_{sc}$ is positive \cite{Pu2003}.
 It is very interesting to
consider the possibility of superfluidity at negative $a_{sc}$ in
a negative-temperature system. One expects the superfluid phase to
allow for dissipationless motion of objects, unless a critical
velocity is exceeded, in which case negative dissipation sets in.
Further properties of negative-temperature superfluidity are
hitherto unexplored theoretically as well as experimentally.

In conclusion, in an optical lattice  combined with a parabolic
potential, a large phase space in which atoms can thermalize to
negative kinetic temperatures is open for exploration.

\acknowledgements{I have greatly benefited from inspiring
 discussions with Claude Cohen-Tannoudji, Ad Lagendijk and Willem
 Vos.
}


\begin{thebibliography}{99}

\bibitem{Einstein1905} A. Einstein, Ann. Physik (Leipzig) \textbf{17}, 549
(1905). 
\bibitem{Kubo1966} R Kubo, Rep. Prog. Phys. \textbf{29}, 255-284
(1966).

\bibitem{Strandberg1957} M. W. P. Strandberg, Phys. Rev. \textbf{106}, 617
(1957).

\bibitem{Geusic1967} J.E. Geusic, E. O. Schultz-Dubois, and H. E.
D. Scovil, Phys. Rev \textbf{156}, 343 (1967).

\bibitem{Nakagomi1980} T. Nakagomi, J. Phys. A:Math. Gen.
\textbf{13}, 291 (1980).

\bibitem{Hsu1992} W. Hsu and R. Barakat, Phys. Rev. B \textbf{46},
6760 (1992).

\bibitem{Purcell1951} E. M. Purcell and R. V. Pound, Phys. Rev. \textbf{81}, 279 (1951).
\bibitem{Ramsey1956} N. F. Ramsey, Phys. Rev. \textbf{103}, 20 (1956).
\bibitem{Klein1956} M. J. Klein, Phys. Rev. \textbf{104}, 589
(1956).

\bibitem{LandauLifshitz} L. D. Landau and E. M. Lifshitz, \textit{Statistical Physics},
$3^{rd}$ ed. (Pergamon, New York, 1980), p. 221.

\bibitem{KittelKroemer} C. Kittel and H. Kroemer, \textit{Thermal Physics}, $2^{nd}$ ed. (Freeman, San Francisco,
1980), appendix E.

\bibitem{Jessen1996} P. S. Jessen and I. H. Deutsch, Adv. At. Mol.
Phys. \textbf{37}, 95 (1996).

\bibitem{Dahan1996} M. Ben Dahan, E. Peik, J. Reichel, Y. Castin,
and C. Salomon, Phys. Rev. Lett. \textbf{76}, 4508 (1996).

\bibitem{Eiermann2003} B. Eiermann \textit{et al.}, Phys. Rev.
Lett \textbf{91}, 060402 (2003).


\bibitem{Pu2003} H. Pu, L. O. Baksmaty, W. Zhang, N. P. Bigelow,
and P. Meystre, Phys. Rev. A \textbf{67}, 043605 (2003).

\bibitem{Gradshteyn} I. S. Gradshteyn and I. M. Ryzhik, Table of
Integrals, Series, and Products, 5th edition (Academic Press, San
Diego, 1994).

\bibitem{Bloch2002}
M. Greiner, O. Mandel, T. Esslinger, T.W. H\"{a}nsch, and I.
Bloch, Nature (London) \textbf{415}, 39-44 (2002).


\bibitem{Hooley2004prl} C. Hooley and J. Quintanilla, Phys. Rev. Lett. \textbf{93}, 080404 (2004).


\bibitem{Zener1934} C. Zener, Proc. R. Soc. London Ser. A \textbf{145},
523 (1934).


\bibitem{Zheng2004} Y. Zheng, M. Ko\v{s}trun, and J. Javanainen,
Phys. Rev. Lett. \textbf{93}, 230401 (2004).

\bibitem{Brand2005} J. Brand and A. Kolovsky, cond-mat/0412549 (2004).

\bibitem{Inouye1998} S. Inouye \textit{et al.}, Nature
\textbf{392}, 151-154 (1998).

\bibitem{Luiten1996} O.J. Luiten, M.W. Reynolds, and J.T.M. Walraven,
 Phys. Rev. A.
\textbf{53}, 381 (1996).

\end{thebibliography}
\end{document}